\documentclass[aoas,MSNbibl,nameyear,dvips]{arximspdf}
\usepackage{graphicx}

\doi{10.1214/11-AOAS492}
\volume{5}
\issue{4}
\pubyear{2011}
\firstpage{2385}
\lastpage{2402}



\begin{document}
\begin{frontmatter}

\title{Multivariate Bayesian semiparametric
models for authentication of food and beverages\thanksref{T1}}
\runtitle{Multivariate Bayesian semiparametric models}

\thankstext{T1}{The anthocyanin profiles were obtained in the frame of the FONDEF Grant
D00I1138.}

\begin{aug}
\author[A]{\fnms{Luis} \snm{Guti\'errez}\corref{}\ead[label=e1]{luisgutierrez@med.uchile.cl}}
\and
\author[B]{\fnms{Fernando A.} \snm{Quintana}\thanksref{t2}\ead[label=e2]{quintana@mat.puc.cl}}
\runauthor{L. Guti\'errez and F. A. Quintana}
\affiliation{Universidad de Chile and Pontificia Universidad Cat\'olica
de Chile}
\address[A]{Divisi\'{o}n de Bioestad\'{i}stica\\
Escuela de Salud P\'{u}blica\\
Facultad de Medicina\\
Universidad de Chile\\
Avenida Independencia 939, Santiago\\
Chile\\
\printead{e1}} 
\address[B]{Departamento de Estad\'istica\\
Facultad de Matem\'aticas\\
Pontificia Universidad Cat\'olica de Chile\\
Casilla 306, Correo 22, Santiago\\
Chile\\
\printead{e2}}
\end{aug}

\thankstext{t2}{Supported in part by Grant FONDECYT 1100010.}

\received{\smonth{11} \syear{2010}}
\revised{\smonth{6} \syear{2011}}

%
\begin{abstract}
Food and beverage authentication is the process by which foods or
beverages are verified as complying with its label description, for example,
verifying if the denomination of origin of an olive oil bottle is
correct or if the variety of a certain bottle of wine matches its label
description. The common way to deal with an authentication process is
to measure a number of attributes on samples of food and then use these
as input for a classification problem. Our motivation stems from data
consisting of measurements of nine chemical compounds denominated
Anthocyanins, obtained from samples of Chilean red wines of grape
varieties Cabernet Sauvignon, Merlot and Carm\'{e}n\`{e}re. We consider
a model-based approach to authentication through a semiparametric
multivariate hierarchical linear mixed model for the mean responses,
and covariance matrices that are specific to the classification
categories. Specifically, we propose a model of the ANOVA-DDP type,
which takes advantage of the fact that the available covariates are
discrete in nature. The results suggest that the model performs well
compared to other parametric alternatives. This is also corroborated by
application to simulated data.
\end{abstract}

%

\end{frontmatter}

\section{Introduction}\label{sectintr}

Food and beverage authentication is the process in\break which foods or
beverages are verified as complying with its label description
[Winterhalter (\citeyear{Winterhalter07})]. From the viewpoint of consumers' acquisition,
the mislabeling of foods represents commercial fraud [\citet{Mafra08}].
On the other hand, producers and sellers could have problems if their
products are mislabeled. Food authentication is important for foods and
beverages of high commercial value, like honey, wines or olive oil,
because their prices depend of their quality, variety or origin. It is
then important to uncover unscrupulous sellers who decide to increase
their profit by adulterating these products with similar but lower
quality substances. Misleading labeling might also have negative health
implications, especially when the food has undeclared allergenic
compounds.

Because of the growing demand from consumers of clarity and certainty
in food origins and contents, the importance of food authentication has
substantially increased in recent years. Many analytical tools and
methods used for authenticity have been consequently developed. In
particular, there is a very active area of research on the
determination of chemical markers for classification and/or
authentication of wines. Anthocyanin profiles are known to be specially
useful for the purpose of wine variety authentication. See, for example,
\citet{eder94}, \citet{Berente00}, \citet{Holbach01},
\citet{Revilla01}, \citet{Otteneder04} and \citet{vonbaer07}.

Data analysis methods for authentication purposes have been developed
mainly outside the statistics fields, and most of them are exploratory
techniques designed to deal with multivariate data sets. Probabilistic
modeling for discrimination and authentication purposes was proposed by
\citet{Brown99}, who used Bayesian methods to discriminate $39$
microbiological taxa using their reflectance spectra. More recently,
\citet{Deanetal06} used a Gaussian mixture model with labeled and
unlabeled samples, with application to the authentication of meat
samples from five species, and the geographic origin of olive oils.
\citet{Toher07} compared model-based classification methods such as
Gaussian mixtures with partial least squares discriminant analysis,
considering samples of pure and adulterated honey.

We propose a model-based procedure to solve the authentication problem
of foods and beverages. The motivation comes from a data set consisting
of measurements of nine chemical compounds denominated Anthocyanins,
obtained from samples of Chilean red wines of grape varieties Cabernet
Sauvignon, Merlot and Carm\'{e}n\`{e}re. We propose a semiparametric
Bayesian model that allows us to define a flexible distribution $G$ for
the joint measurements. The model has the advantage of not having to
assume any parametric form, which may be particularly difficult to
check in multivariate cases. Increased flexibility is added by allowing
$G$ to be formulated under the formalism of dependent random
probability measures as in \citet{Deiorio04}. An interesting aspect
of the proposed approach is the need to extend previous univariate
semiparametric models as in \citet{delacruzb07} to the multivariate
case.

The rest of the paper is organized as follows. We first present the
wine data set and the related authentication problem in
Section \ref{sectmotivating}. In Section~\ref{secttheo} we give a
brief theoretical background about Bayesian semiparametric models and
dependent Dirichlet processes, and discuss our approach to the
authentication problem. In Section \ref{sectmodel} we present the
model, which is an extension of the univariate semiparametric Bayesian
linear mixed model [\citet{dey1998practical}] to the multivariate case.
In Section \ref{sectperformance} we illustrate the performance of the
proposed model in a simulated data set. In Section \ref{sectwinedata}
we apply the model to authenticate red wines samples based on their
anthocyanin profile. The paper concludes in Section \ref{sectdisc}
with a discussion and final remarks.

\section{The motivating data set}\label{sectmotivating}

We consider a data set consisting of measurements of concentrations of
nine anthocyanins on samples of Chilean red wines. Anthocyanins are a
group of chemical compounds present in red wine, which confer to this
beverage its characteristic red color and are transferred from the
grape skins to wine during the winemaking process. The data set includes
the grape variety for each sample \textit{as declared by the
producer}, the year of harvest and the geographical origin or valley.
The grape varieties in the data set are Cabernet Sauvignon (228
samples), Carm\'en\`ere (95 samples) and Merlot (76 samples). These
samples form a~data set with mixed wine types, which represents the
most abundant red grape varieties cultivated in Chile across different
valleys. The data are unbalanced, and there are combinations of variety
and location for which no observations are available.
All wine samples came directly
from wineries located in the valleys of Aconcagua, Maipo, Rapel,
Curic\'{o}, Maule, Itata and B\'{\i}o-B\'{\i}o in Chile. They
correspond to the
vintages 2001, 2002, 2003 and 2004. Anthocyanin determination was made
by reverse phase High Performance Liquid Chromatography (HPLC), a
Chromatographic technique that can separate a mixture of compounds and
is used in analytical chemistry to identify, quantify and purify the
individual components of complex mixtures, like wines and others
beverages or foods. The analytical chemistry procedure was based on
the method described by \citet{Holbach97}, \citet
{Otteneder02} and by
the International Organization of Vine and Wine (OIV) [\citet{OIV03}],
with some minor modifications. More details about anthocyanin
determination for this data set can be found in von Baer et~al.
(\citeyear{vonbaer05,vonbaer07}). A main concern for these data is the
authentication of grape variety using the log-concentrations of the
following anthocyanins: delphinidin-3-glucoside (DP),
cyanidin-3-glucoside (CY), petunidin-3-glucoside (PT),
peonidin-3-glucoside (PE), malvidin-3-glucoside (MV),
peonidin-3-acetylglucoside (PEAC), malvidin-3-acetylglucoside\break (MVAC),
peonidin-3-coumaroylglucoside (PECU) and
malvidin-3-couma\-roylglucoside
(MVCU). To do so, we propose in Section \ref{sectmodel} a
multivariate linear mixed model that attempts to characterize the
variability in anthocyanin log-concentrations in terms of variety and
valley of origin. We also point out that we ignore vintage year
in our development. The pragmatical reason for this is that by doing so
we may easily incorporate data from new years as they become available,
without the need to modify the model. In support of this choice, we
refer to \citet{Gutietal10} who used the year of harvest as a
continuous predictor when proposing a Bayesian parametric model for the
same data. The idea was to overcome this very same limitation. Yet, the
effect of vintage year was negligible in that context.

\section{Some background material}\label{secttheo}

Semiparametric models have both parametric and nonparametric parts,
the distinction between these being that the parameters belong to a
finite and infinite-dimensional space, respectively. Semi- and
nonparametric Bayesian models are used mainly to avoid critical
dependence on parametric assumptions. An important application of such
modeling strategy is to random effects distributions in
hierarchical models, where often little is known about the specific
form of such distributions [\citet{Muller04}]. To handle the
nonparametric part of the model, we need to define a random measure on
the space of distribution functions. The most popular such choice is
the Dirichlet process (DP) [\citet{Ferguson73}].

In a food authentication context scenario, we need to build a model
that adequately accounts for all the problem-specific features. In
the specific case of our motivating data set, it is reasonable to
think of wines coming from the same valley as being correlated, because
soil and weather conditions are similar within a given valley. The
usual (and simplest) way to induce a~correlation structure is by
incorporating random effects or sample specific parameters in a~model.
Let $\alpha_i$ denote the random effects and let $z_i$ be a~categorical
covariate with $k$ levels (e.g., $k$ different regions of origin). We
could assume a single nonparametric prior on $\alpha_i$ for all
samples, without reference to the levels of $z_i$. Alternatively, we
could consider differences by putting $k$ independent priors on
$\alpha_i$. These two extreme modeling strategies imply that
$G_{z_1}=\cdots=G_{z_k}$ for the former and $G_{z_1},\ldots,G_{z_k}$ to
be mutually independent for the latter. \citet{MacEachern99} proposes
a modeling strategy, the Dependent Dirichlet Processes (DDP), that
allows the set of random effects distributions to be similar but not
identical to each other. \citet{MacEachern99} defines a nonparametric
probability model for $G_z$ in such a way that marginally, for each
$z=z_{j}$ $(j=1,\ldots,k)$, the random measure $G_z$ follows a DP. In
this context, the DP representation proposed by \citet{Sethuraman94}
is quite useful. Sethuraman's representation establishes that any
$G\sim \operatorname{DP}(M,G_0)$ can be represented as an infinite mixture of point
masses:\looseness=-1
%
\begin{eqnarray}
G(\cdot) &=& \sum_{h=1}^\infty w_h\delta_{\mu_h}(\cdot),\qquad
\mu_h\stackrel{\mathrm{i.i.d.}}\sim G_0,\nonumber\\[-8pt]\\[-8pt]
w_h &=& U_h \prod_{j<h} (1-U_j)  \qquad\mbox{with }   U_h
\stackrel{\mathrm{i.i.d.}}\sim \operatorname{Beta}(1,M).\nonumber
\end{eqnarray}
The key idea behind the DDP is to introduce dependence across the $G_z$
measures by assuming the\vadjust{\goodbreak} distributions of the point masses to be
dependent across different levels of $z$ (i.e., $\mu_{zh}$), but still
independent across $h$. If the weights are assumed to be the same
across $z$, the dependent probability measure can be represented as
$G_z(\cdot) = \sum_{h=1}^\infty w_h\delta_{\mu_{zh}}$. The last idea
was used by \citet{Deiorio04} in the construction of an ANOVA-DDP type
model. A similar approach was used in spatial modeling by
\citet{Gelfand05}, who proposed a Gaussian process for the atoms,
\citet{Caron06} in times series, \citet{delacruzb07} in
classification, \citet{deiorioetal09} in survival analysis and,
recently, by \citet{Jara10} who proposed a Poisson--Dirichlet process
for the analysis of a data set coming from a dental longitudinal study.
\citet{Griffin06} point out that letting only the atoms depend on
covariate values may lead to certain problems when points in the domain
are far from the observed data. They propose an approach that avoids
this by locally updating the process and inducing dependence in the
weights through distance-based similarities in the ordering of atoms,
through viewing the atoms as marks in a point process. Other works
where covariate dependence is introduced in the weights are
\citet{Dunson07} and \citet{Dunson08}. \citet
{Muller96} considered
a completely different approach for inducing dependence in $G$. They
used a DP mixture of normals for the joint distribution of $y$ and $z$,
and then focused on the implied conditional density of $y$ given $z$
for estimating the mean regression function. A recent reference about
nonparametric Bayesian statistics, DDP models and their applications
can be found in \citet{Hjort10}.

The almost sure discreteness of the Dirichlet process makes it
inappropriate as a model for a continuous quantity $y$. A standard
procedure for overcoming this difficulty is to introduce an additional
convolution so that
%
\begin{equation}\label{mixture}
H(y) = \int f(y\mid\theta)\,dG(\theta)  \qquad\mbox{with }
G\sim \operatorname{DP}(M,G_0).
\end{equation}
Such models are known as DP mixtures (DPM) [\citet{Antoniak74}]. The
mixture model (\ref{mixture}) can be equivalently written as a
hierarchical model by introducing latent variables $\theta_i$ and
breaking the mixture as
%
\begin{equation}
y_i\mid\theta_i \sim f(y_i\mid\theta_i), \qquad \theta_i\sim G
\quad \mbox{and} \quad  G\sim \operatorname{DP}(M, G_0).
\end{equation}

For the majority of food authentication problems the responses are
continuous multivariate and covariates are discrete. This is the case
for the data described in Section \ref{sectmotivating}. Thus, we will
adopt the popular semiparametric modeling strategy that consists of
introducing dependence in the random effects distribution and then
adding a convolution with a continuous kernel. The ANOVA-DDP approach
of \citet{Deiorio04} is a natural way to build the desired dependence
into the model, as will be discussed below in Section \ref{sectmodel}.
We remark here that a model that defines dependence in terms of
distances would not be appropriate for an authentication problem with
categorical covariates, as is our case.

\section{The model} \label{sectmodel}

We first note that due to the multivariate nature of many
authentication problems (which is also the case of the wine data), it
would not be appropriate to treat the individual responses in an
univariate way.

We assume that the $i$th response vector is related to the
covariates in a~linear way. Furthermore, we assume that there are fixed
and random effects in the model. The model for the $i$th unit in
the $u$th group or class in our classification context is
thus assumed to be given by
%
\begin{eqnarray}\label{mixturemodel}
(y_{iu}\mid x_{iu},z_{iu}) &\sim& N_p(Bx_{iu}+\theta_{iu}, \Sigma_u),
\qquad
i=1,\ldots,n_u,  u=1,\ldots, m, \nonumber\\
\theta_{iu}&\sim& H_z(\theta_{iu}),\nonumber\\[-8pt]\\[-8pt]
H_z(\theta) &=& \int N(\theta\mid z\alpha, \tau)\,dG(\alpha) ,\nonumber\\
G &\sim& \operatorname{DP}(M,G_0), \nonumber
\end{eqnarray}
where $y_{iu}$ is a vector of responses in $R^p$, $B$ is a $p\times q$
matrix of fixed effects,~$x_{iu}$ is a vector of covariates in $R^q$,
$\theta_{iu}$ is a $p\times1$ vector of unit-specific random effects,
$z_{iu}$ is a $p\times pk$ design matrix for random effects and
$\alpha$ is a $pk\times1$ vector of latent variables that define the
random effects. For classification purposes, the subscript $u$
denotes the group or class. Model (\ref{mixturemodel}) implies that
$H_z(\theta) = \sum_{h=1}^\infty w_h N(\theta\mid z\alpha_h,\tau)$ is
an infinite mixture of normal distributions. As usual in mixture
models, posterior simulation proceeds by breaking the mixture in
(\ref{mixturemodel}) by introducing latent variables~$\alpha_i$:
%
\begin{equation}\label{eqmainmodel}
\theta_{iu} = z_{iu}\alpha_i+\eta_i,\qquad \alpha_i\sim G,\qquad
G\sim \operatorname{DP}(M,G_0)\quad \mbox{and}\quad  \eta_i\sim
N_p(0,\tau).\hspace*{-25pt}
\end{equation}
By simplicity, we choose a multivariate normal model for the base
measure $G_0\equiv N_{pk}(0,R)$ and as usual in this context, we assume
prior independence for all remaining parameters. The prior distribution
for matrix $B=[\beta_1,\beta_2,\ldots,\beta_q]$ is assumed to be
independent by columns, that is, $\beta_1,\beta_2,\ldots,\beta_q$ are
mutually independent with distribution given by
%
\begin{equation}\label{distbeta}
\beta_1,\ldots,\beta_q \sim N_p(\beta_{0j},\Lambda),\qquad
j=1,\ldots,q.
\end{equation}
The prior distributions for the variance--covariance matrices
$\Sigma_u$, $u=1,\ldots,m$, and $\tau$ are given by
%
\begin{equation}\label{distsigmatau}
\Sigma_1,\ldots,\Sigma_m \sim IW_p(\nu_0,Q_0),\qquad
\tau\sim IW_{p}(\gamma_0, \Phi_0).
\end{equation}
We complete the Bayesian formulation of model (\ref{mixturemodel}) by
specifying the prior for hyperparameters $R$,
$\beta_{01},\ldots,\beta_{0q}$, $\Lambda$ and $M$ as
%
\begin{eqnarray}\label{distothers}
R &\sim& IW_{pk}(r_0, R_0),\qquad \beta_{01},\ldots,\beta_{0q}
\sim N_p(\alpha_0,\tau_0),\\
\Lambda&\sim& IW_p(L_0,t_0),\qquad M \sim Ga(a_1, a_2).
\end{eqnarray}

The random distribution $H_z(\theta)$ in model (\ref{mixturemodel}) is
dependent of the level of covariate~$z$. As such, this follows the
model proposed by \citet{Deiorio04}, but our focus is on the
application to multivariate data. Matrix~$R$ in the model allows for
correlation between all components of vector~$\alpha_i$, which implies
correlation between different components of the response vector and
between different levels of~$z$. The full conditional posterior
distributions and details of the posterior simulation scheme are given
in the \hyperref[app]{Appendix}.\looseness=1

Consider now a general classification approach, and denote the
training data set by $y^n=(y_{1},\ldots,y_{n}, x_{1},\ldots,x_{n},
z_{1},\ldots,z_{n}, g_{1},\ldots,g_{n})$. Here, $y_{iu}=(y_i\dvtx g_i=u)$,
$u=1,\ldots,m$, is the response vector for the $u$th group,
$x_{iu}=(x_i\dvtx g_i=u )$ is the vector of covariates for fixed effects,
$z_{iu}=(z_i\dvtx g_i=u)$ is a~vector of covariates for random effects and
$g_{i}$ represents the known group label for the $i$th unit.
Consider a new unit for which the response $y_{n+1}$ and covariate
vectors $x_{n+1}$ and $z_{n+1}$ are known, but its label $g_{n+1}$ is
unknown. We want to assign a label $u$ to the new unit, where $u \in
\{1,\ldots,m\}$. Consequently, it is necessary to estimate the
classification probability $P(g_{n+1}=u\mid y_{n+1},y^n)$. Following
\citet{delacruz07} and \citet{Gutietal10}, we use
%
\begin{equation}\label{eqapproxChapter3}
P(g_{n+1}=u\mid y_{n+1},y^n)\approx
\frac{1}{C}\sum_{c=1}^{C}\frac{\pi_up(y_{n+1}\mid\Theta_u^{(c)})}{\sum
_l\pi_lp(y_{n+1}\mid\Theta_l^{(c)})}.
\end{equation}
In (\ref{eqapproxChapter3}), $\pi_u=P(g_i=u)$ may be taken as the
empirical group proportions. We propose classifying an existing unit,
$i$, and a future one, $n+1$, using the zero--one law considered in
\citet{Hastie01}:
%
\begin{eqnarray}\label{eqclassification}
\hat{g_i}&=&\mathop{\arg\max}_{u} P(g_i=u\mid y^n)
\quad\mbox{and}\nonumber\\[-8pt]\\[-8pt]
\hat{g}_{n+1}&=&\mathop{\arg\max}_u P(g_{n+1}=u\mid y^n,y_{n+1}),\nonumber
\end{eqnarray}
that is, assigning the label as the category that maximizes the
classification probability (\ref{eqapproxChapter3}). Note that
(\ref{eqapproxChapter3}) and (\ref{eqclassification}) are part of a
generic classification approach which is based on predictive
distributions. In particular, (\ref{eqapproxChapter3}) relates to the
model described in (\ref{mixturemodel}) through the $\Theta_u^{(c)}$
quantities, which represent samples from the posterior distribution for
model (\ref{mixturemodel}), given the training data set $y^n$. For the
wine data analysis later in Section \ref{sectwinedata}, we will let
the fixed effects be varieties and random effects be the different
regions of origin. For classification purposes, we are assuming that a
new wine sample is available, for which we know its anthocyanin
concentrations and valley, and the aim is to predict its variety. Note
that the variety here corresponds to the label $u$ and there are not
other covariates for fixed effects, so the rows on the matrix of a
fixed effects are given by $x_i=(1,0,0)$ if $g_i=1$ (Cabernet
Sauvignon), $x_i=(0,1,0)$ if $g_i=2$ (Merlot) and $x_i=(0,0,1)$ if
$g_i=3$ (Carm\'en\`ere).

\section{Classification performance of the proposed model}
\label{sectperformance}

\begin{figure}

\includegraphics{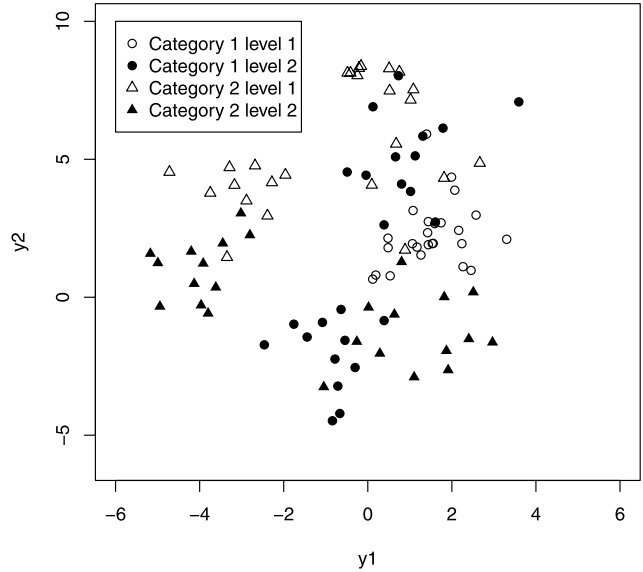}

\caption{Simulated data set.}\label{figsimulated}
\end{figure}

To evaluate the classification performance of the proposed model, we
simulated a data set considering $m=2$, $n=100$, $p=2$, $q=2$, $k=2$. We
simulated from a mixture of $p$-variate normal distributions,
$\sum_{i=1}^8\omega_iN(\mu_i,\Sigma)$,\vspace*{1pt} where $\omega_1,\ldots,\omega_8$
are given by (0.25, 0.12, 0.13, 0.1, 0.1, 0.05, 0.12, 0.13),
respectively, $\mu_1=(1.1,2.3)^t$, $\mu_2=(0.1,-2)^t$,
$\mu_3=(1.3,5)^t$, $\mu_4=(-3,3.4)^t$, $\mu_5=(-0.1,7)^t$,
$\mu_6=(1.8,5)^t$, $\mu_7=(-4,1)^t$, $\mu_8=(1,-2)^t$ and $\Sigma$ is
given by $\sigma_{11}=0.932$, $\sigma_{12}=0.11$ and
$\sigma_{22}=1.632$. Figure \ref{figsimulated} shows the simulated
data set. Here, $m=2$ means that we have to classify between two
categories with $k=2$ levels for the covariate $z$.
%
The hyperparameter values were taken as $\alpha_0=(0,0)^t$,
$\tau_0=100I_2$, $Q_0=I_2$, $L_0=I_2$, $\nu_0=4$, $r_0=4$, $t_0=4$,
$R_0=I_{pk}$, $\gamma_0=4$, $\phi_0=0.001I_p$ and $a_1=a_2=1$.
Table~\ref{tableclassiperformance} shows the classification results of
%
\begin{table}[b]
\caption{Classification performance for the simulated data set.
Values within parenthesis were obtained using leave-one-out
cross-validation technique} \label{tableclassiperformance}
%
\begin{tabular*}{\tablewidth}{@{\extracolsep{\fill}}lcccccc@{}}
\hline
& \multicolumn{2}{c}{\textbf{BSP}} & \multicolumn{2}{c}{\textbf{BP}} &
\multicolumn{2}{c@{}}{\textbf{LDA}} \\[-4pt]
& \multicolumn{2}{c}{\hrulefill} & \multicolumn{2}{c}{\hrulefill} &
\multicolumn{2}{c@{}}{\hrulefill} \\
\textbf{Category} & \textbf{1} & \textbf{2} & \textbf{1} & \textbf{2}
& \textbf{1} & \textbf{2} \\
\hline
1 & 47 (43) & 3 (7) & 47 (35) & \hphantom{0}3 (15) & 42 (42) & 8 (8) \\
2 & 3 (9) & 47 (41) & 9 (9) & 41 (41) & 17 (19) & 33 (31) \\
\hline
\end{tabular*}
%
\end{table}
the proposed Bayesian semiparametric model (BSP), comparing with linear
discriminant analysis (LDA), which is the usual technique used in the
literature for this type of problem, and a parametric (BP) version of
model~(\ref{mixturemodel}), defined as
%
\begin{eqnarray}\label{parametricmodel}
(y_{iu}\mid x_{iu},z_{iu}) &\sim& N_p(Bx_{iu}+\theta_{iu},
\Sigma_u),\qquad
i=1,\ldots,n,  u=1,\ldots,m ,\nonumber\\
\theta_{iu}&=& z_{iu}\alpha+\eta_i, \qquad \eta_i\sim N_p(0,\tau), \\
\alpha&\sim& N_{pk}(0,R). \nonumber
\end{eqnarray}

\begin{figure}

\includegraphics{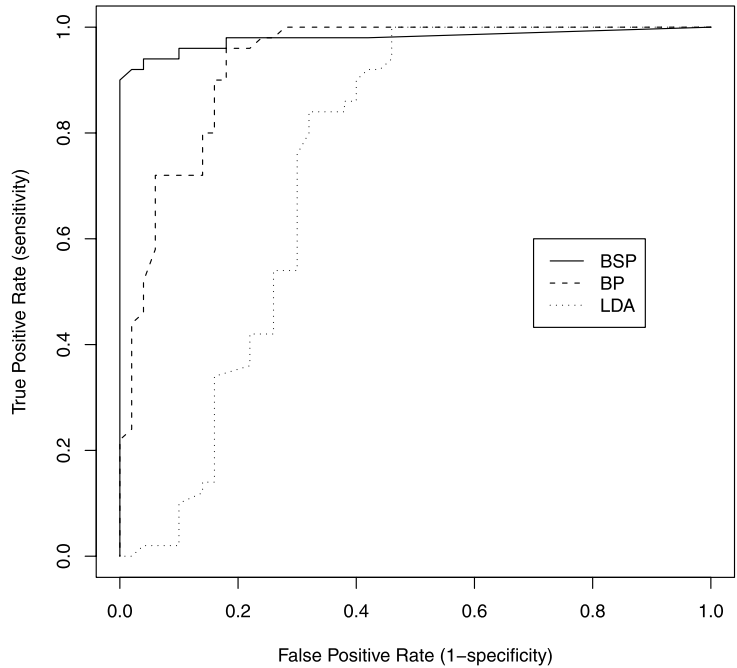}

\caption{ROC curves for classification of the simulated data set under
Bayesian semiparametric model (BSP), Bayesian parametric (BP) and linear
discriminant analysis (LDA).} \label{figROC}
\end{figure}

Using the proposed BSP model, we obtained a classification error of
7.0\% in the training set and 16.0\% using leave-one-out cross-validation
(LOOCV). In contrast, the BP model resulted in a classification error
of 12.0\% in the training set and 24.0\% under LOOCV, while the
corresponding figures for the LDA were 25.0\% and 27.0\%, respectively. A
common way to assess the performance of classification rules is the
Receiver Operating Characteristic curve (ROC) shown in
Figure~\ref{figROC}, which plots the true positive rate against the
false positive rate for all the different possible cutpoints. From the
ROC curves we also calculated the Area Under ROC curve (AUC) for the
three models, with higher values corresponding to models with better
discrimination capabilities. We obtained 0.9792 for the BSP model,
0.9334 for the BP model, and 0.7464 for LDA. These results clearly
suggest the superiority of the proposed BSP model for wine
authentication in our simulation, compared to the other
alternatives.

Another important aspect of the analysis concerns comparing model
adequacy of the BP versus our BSP proposal. To this effect, we
calculated the Conditional Predictive Ordinates ($\mathrm{CPO}_i$)
[\citet{Chen00}], summarized in the log-pseudo marginal likelihood
statistic $\mathrm{LPML}=\break\sum_{i=1}^n \log(\mathrm{CPO}_i)$ [\citet{geissereddy79}].
Models with higher LPML values are to be preferred. We found the
proposed BSP to perform better than the BP, as the corresponding LPML
values were $-365.3$ and $-428.7$, respectively. As an additional
comparison we considered the Deviance Information Criterion (DIC)
[\citet{Spiegelhalter02}]. Models with lower DIC values are to be
preferred. It is important to point out that several possible DIC
construction are available. \citet{celeux06} discuss the DIC for
finite mixture and random effects models, giving and comparing
different DIC definitions. Our proposal is a mixture model, but the
mixture is infinite. Indeed, due to the convolution in
(\ref{mixturemodel}), the random effects $\{\theta_i\}$ have continuous
distributions. Nevertheless, we found some constructions of DIC to
yield negative values for the effective dimension $p_D$, and so we used
$\mathrm{DIC}_1$, $\mathrm{DIC}_2$ and $\mathrm{DIC}_3$ of \citet{celeux06}. The corresponding
values were $721.4$, $717.3$ and $723.1$ for the BSP model, and
$856.5$, $854.4$ and $855.4$ for the BP model, and the conclusion is
the same regardless of the specific construction. In summary, both LPML
and DIC criteria consistently point to the superiority of the BSP model
compared to the BP one. Overall, the results suggest that the BSP model
is more flexible, especially when the data cluster between and within
covariate levels.

\section{Performance of the model with wine data set}\label{sectwinedata}

We consider now application of the proposed BSP model to the wine
data set. The response vector is formed by the nine anthocyanins listed
in Section \ref{sectmotivating}. As covariates, we use grape variety
(fixed effects) and valleys (random effects). The hyperparameter values
were taken as $\alpha_0=(0,0,0,0,0,0,0,0,0)^t$, $\tau_0=100I_9$,
$Q_0=0.1I_9$, $L_0=0.01I_9$, $\nu_0=11$, $r_0=65$, $t_0=11$,
$R_0=10I_{pk}$, $\gamma_0=11$, $\phi_0=0.01I_p$ and $a_1=a_2=1$, where
$p=9$, $q=3$ and $k=7$. The resulting prior densities are proper, but
the one for $B$ is vague and hence relatively uninformative. The prior
density for $R$ is relatively uninformative too. All the prior
covariance matrices were assumed of diagonal form. The selected
hyperparameter values imply proper but vague prior distributions,
representing the lack of genuine prior information on the parameters.
To further investigate how sensitive the results are to the above
choices, we conducted a prior sensitivity analysis for hyperparameters
$(a_1,a_2)$, which control the implied clustering structure, and
$\tau_0$, which controls the prior variance for fixed effects. We tried
the prior settings listed in the leftmost column of
Table \ref{tablepriorsensitivity}, which also shows the posterior
%
\begin{table}
\caption{Posteriors means and standard deviations (S.D.) for some
model quantities. For the combinations indicated in the leftmost
column, we present summaries for $M$, $\tau$ and for the fixed effects
parameters for Cabernet Sauvignon corresponding to anthocyanins
PT~($\beta_{13}$) and MV ($\beta_{15}$)}\label{tablepriorsensitivity}
\begin{tabular*}{\tablewidth}{@{\extracolsep{\fill}}lcrlrl@{}}
\hline
& & \multicolumn{2}{c}{\textbf{PT}}
& \multicolumn{2}{c@{}}{\textbf{MV}} \\[-4pt]
& & \multicolumn{2}{c}{\hrulefill}
& \multicolumn{2}{c@{}}{\hrulefill} \\
\textbf{Prior value} & \multicolumn{1}{c}{\textbf{Precision parameter}}
& \multicolumn{1}{c}{\textbf{Mean}} & \multicolumn{1}{c}{\textbf{(S.D.)}}
& \multicolumn{1}{c}{\textbf{Mean}} & \multicolumn{1}{c}{\textbf{(S.D.)}} \\
\hline
$a_1=a_2=0.01$ & $M=1.41$ (0.62) & $\beta_{13}=2.55$
& (0.14) & $\beta_{15}=4.91$ & (0.12) \\
$\tau_0=100I_9$ & & $\tau=0.15 $ & (0.0136) & $\tau=0.08 $
& (0.0088) \\ [4pt]
$a_1=a_2=1$ & $M=1.65$ (0.57) & $\beta_{13}=2.68$ & (0.09)
& $\beta_{15}=5.07$ & (0.11) \\
$\tau_0=100I_9$ & & $\tau=0.16 $ & (0.0134) & $\tau=0.08 $
& (0.0088) \\ [4pt]
$a_1=1$, $a_2=0.1$ & $M=1.48$ (0.56) & $\beta_{13}=2.76$
& (0.12) & $\beta_{15}=5.03$ & (0.11) \\
$\tau_0=100I_9$ & & $\tau=0.16 $ & (0.0129) & $\tau=0.08 $
& (0.0081) \\ [4pt]
$a_1=10$, $a_2=1$ & $M=3.27$ (0.86) & $\beta_{13}=2.58$ & (0.11) & $\beta_{15}=4.95$
& (0.09) \\
$\tau_0=100I_9$ & & $\tau=0.16 $ & (0.0126) & $\tau=0.08 $
& (0.0079) \\ [4pt]
$a_1=a_2=1$ & $M=1.43$ (0.50) & $\beta_{13}=2.46$ & (0.17)
& $\beta_{15}=4.93$ & (0.15) \\
$\tau_0=I_9$ & & $\tau=0.16 $ & (0.0132) & $\tau=0.08 $
& (0.0105) \\ [4pt]
$a_1=a_2=1$ & $M=1.21$ (0.48) & $\beta_{13}=2.58$ & (0.12)
& $\beta_{15}=4.89$  & (0.10) \\
$\tau_0=10I_9$ & & $\tau=0.16 $ & (0.0129) & $\tau=0.08 $
& (0.0081) \\ [4pt]
$a_1=a_2=1$ & $M=1.82$ (0.61) & $\beta_{13}=2.69$ & (0.08)
& $\beta_{15}=5.05$ & (0.08) \\
$\tau_0=1000I_9$ & & $\tau=0.17 $ & (0.0134) & $\tau=0.09 $
& (0.0093) \\ [4pt]
$a_1=a_2=1$ & $M=1.38$ (0.52) & $\beta_{13}=2.57$ & (0.14)
& $\beta_{15}=5.0$\hphantom{0} & (0.07) \\
$\tau_0=10000I_9$ & & $\tau=0.16 $ & (0.0125) & $\tau=0.08
$ & (0.0082) \\ \hline
\end{tabular*}
%
%
\end{table}
means and standard deviations for some key parameters in the model.
Specifically, we show posterior summaries for $M$, $\tau$ and for the fixed
effects parameters for Cabernet Sauvignon corresponding to anthocyanins
PT and
MV. We generally found no notorious changes
in these summaries across different prior configurations. Since our
emphasis is on classification, we also compared the shape of the
predictive distributions and the classification itself for the prior
configurations of Table \ref{tablepriorsensitivity}. In general, we
did not observe big changes in the predictive distributions (data not
shown) and the classification was the same for all prior
specifications.

Table \ref{tablemissc} shows the classification results, where the
values within parenthesis were obtained using a LOOCV approach. The
classification error obtained in the training set was 0.5\%, and 3.2\%
under LOOCV. These values are better than those obtained by
\citet{Gutietal10} with the same data set but applying a Bayesian
parametric model, namely, 3.0\% in the training set and 3.5\% using
LOOCV.

%
\begin{table}
\caption{Misclassification rate for the three grape varieties}
\label{tablemissc}
\begin{tabular*}{\tablewidth}{@{\extracolsep{\fill}}lcccc@{}}
\hline
\textbf{Variety} & \textbf{Carm\'en\`ere} & \textbf{C. Sauvignon}
& \textbf{Merlot} & \textbf{Error} \\
\hline
Carm\'en\`ere & 94 (93) & 1 (1) & 0 (1) & 1.1\% (2.1\%)\hphantom{0}\\
C. Sauvignon & 0 (0) & 228 (225) & 0 (3) & 0.0\% (1.3\%)\hphantom{0}\\
Merlot & 1 (8) & 0 (0) & 75 (68) & 1.3\% (10.5\%)\\
[4pt]
Total error & & & & 0.5\% (3.2\%)\hphantom{0}\\
\hline
\end{tabular*}
\vspace*{-6pt}
\end{table}

Table \ref{tableAUC} shows the AUC values, which were calculated based
on separate ROC curves for each grape variety, and for each of the BSP
and BP models. All these values are very high, with the BSP model
attaining the best performance across the three grape varieties.
Comparing the BSP and BP models, we found that the $\mathrm{DIC}_1$, $\mathrm{DIC}_2$,
$\mathrm{DIC}_3$, and LPML statistics values were $-5\mbox{,}901.5$, $-6\mbox{,}368.9$,
$-5\mbox{,}769.9$ and $2\mbox{,}430.1$ for the former, and $-3\mbox{,}659.1$, $-5\mbox{,}006.6$,
$-3\mbox{,}446.2$ and $1\mbox{,}351.6$ for the latter. Again, these results suggest
that the proposed BSP model provides a superior fit, and all the
criteria values are consistent. We also note that in all cases, the
effective dimension~$p_D$ for the DIC was positive.

\begin{table}
\tablewidth=275pt
\caption{Area under ROC curve}\label{tableAUC}
\begin{tabular*}{\tablewidth}{@{\extracolsep{\fill}}lcc@{}}
\hline
\textbf{Grape variety} & \textbf{AUC BSM} & \textbf{AUC BPM} \\
\hline
Cavernet Sauvignon & 0.999999 & 0.9969221 \\
Merlot & 0.999999 & 0.9967403 \\
Carm\'en\`ere & 0.999999 & 0.9963574 \\
\hline
\end{tabular*}
\end{table}

\begin{figure}

\includegraphics{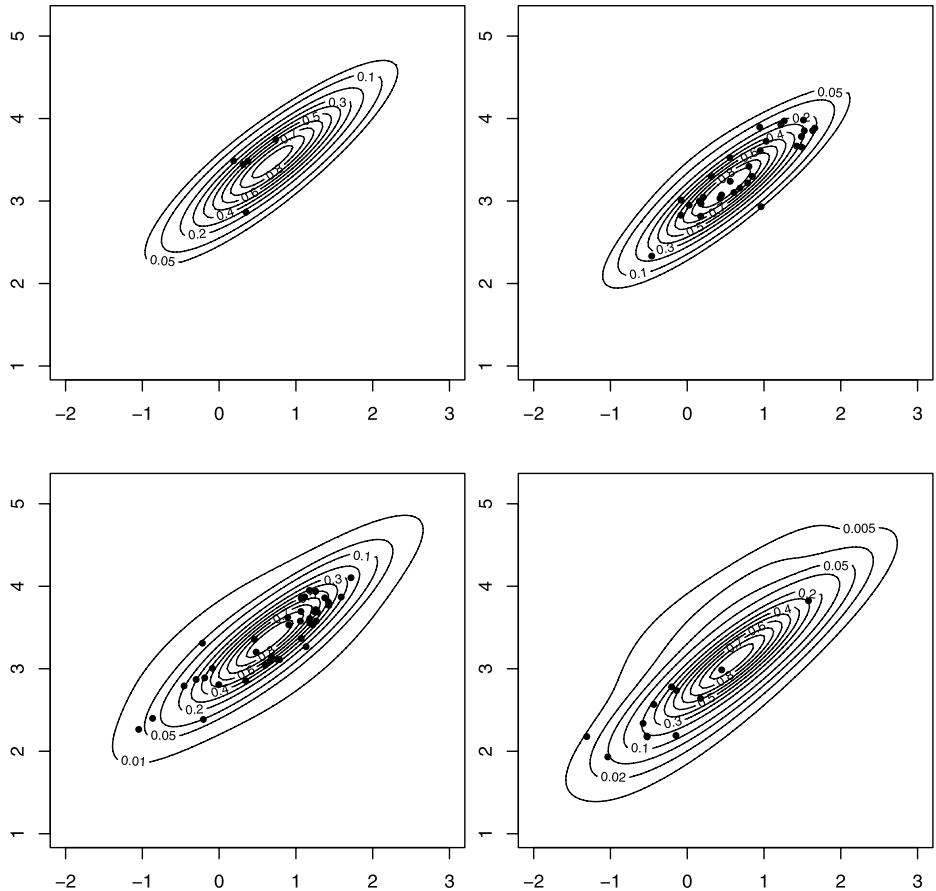}

\caption{Bivariate posterior predictive distributions under the BSP model
for Carm\'en\`ere wines from the Aconcagua, Maipo, Rapel and Curic\'{o}
valleys, with points representing observed values. The anthocyanins
considered here were PECU and MVCU in the horizontal and vertical axes,
respectively.} \label{figpredictive}
\end{figure}

Figure \ref{figpredictive} displays bivariate posterior predictive
distributions for Carm\'en\`e\-re wines from the valleys of Aconcagua,
Maipo, Rapel and Curic\'{o}, considering the PECU and MVCU
anthocyanins. The points on the graph are the observed values. We can
see the changes in the posterior predictive distribution across
valleys. Predictions for the Aconcagua valley show less variation
compared to the Maipo valley. Predictions for The Rapel valley show more
variability, with some evidence of asymmetry, as dictated by the
observed data, but the model provides a reasonable fit to this
behavior. Finally, the Curic\'{o} valley also exhibits asymmetry.

\begin{figure}

\includegraphics{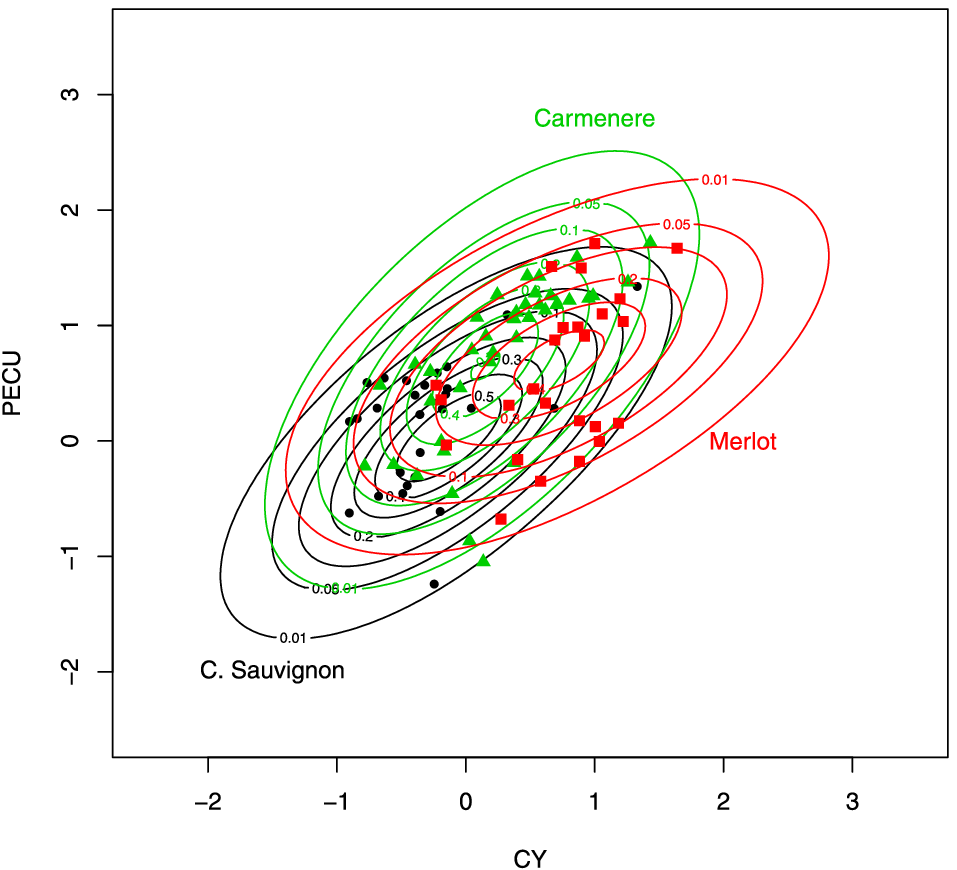}

\caption{Bivariate posterior predictive distributions for Cabernet
Sauvignon, Merlot and Carm\'en\`ere wines from the Curic\'{o} valley,
with points representing the observed values.}
\label{figpredictivethreegrapes}
\end{figure}

Figure \ref{figpredictivethreegrapes} shows the bivariate predictive
posterior distributions for Cabernet Sauvignon, Carm\'en\`ere and
Merlot from the Curic\'{o} valley considering the PECU and CY anthocyanins.
This plot is interesting because it shows how informative are PECU and
CY in terms of the target classification. These two anthocyanins show
%
that some Merlot and Carm\'en\`ere samples have very similar
chemical profiles. This behavior is reasonable because some years
ago, Carm\'en\`ere, which in other countries disappeared due to
phylloxera, was rediscovered in Chile. Formerly, all vineyards planted
with this grape variety in Chile\vadjust{\goodbreak} were declared as Merlot. Using SSR DNA
markers to confirm varietal identity, \citet{Hinrichsen01} found that
from a total of 93 vines of five Chilean vineyards, originally planted
as Merlot, four vines matched Carm\'en\`ere. This leads to the
conclusion that at the time of collecting wine samples, those vineyards
declared as Carm\'en\`ere are correctly identified with high
probability, but a certain percentage of vineyards declared as Merlot
still correspond to Carm\'en\`ere.

\section{Concluding remarks}\label{sectdisc}

We have proposed a linear mixed effects model for wine authentication,
featuring a flexible model for random effects that does not require
to restrict ourselves to a given parametric form. We did so by
resorting to Dependent Dirichlet Processes, which allow the set of
random effects distributions to be similar but not identical to each
other, depending on values of a set of covariates. For the
authentication problem, dependence on covariate levels is important
because it is reasonable to think that foods or beverages that come
either from the same region of origin, or those which were made with
the same technology, could be similar or correlated. The ANOVA-DDP
approach was suitable to our purposes, but other types of nonparametric
priors could be considered.

The proposed BSP model provided a better fit to the data than a
parametric alternative, as we showed in the simulation example and in
the application to the wine data. In terms of the target
classification, the BSP model also provided slightly better results
than other alternatives. Our proposal was motivated by food
authentication, but it could be used in any situation where the aim is
to classify subjects or units into several groups, on the basis
of multiple responses and covariates.

\begin{appendix}\label{app}
\section*{Appendix}

In this section we give the MCMC algorithm that was used for posterior
simulation under the proposed model. Because the model is of conjugate
type, we use algorithm 2 in \citet{Neal00}. Let
$\mathbf{c}=(c_1,\ldots,c_n)$ denote a vector that captures the
clustering of $\alpha_i$ and let $\alpha=(\alpha_c\dvtx c\in
\{c_1,\ldots,c_n\})$. To resample the configurations~$c_i$, we proceed
with the following two steps:

\subsection*{Step 1}

If $c=c_j$ for some $j\neq i$, we compute the probability that the $i$th
element in $\mathbf{c}$ equals other element in the same set as
%
\begin{eqnarray}\label{eqp1}
&&
P(c_i=c\mid c_{-i},\theta_i,\alpha) \nonumber\\
&&\qquad=
b\frac{n_{-i,c}}{n-1+M}(2\pi)^{-p/2}|\tau|^{-1/2}\\
&&\qquad\quad{}\times
\exp\biggl\{-\frac{1}{2}(\theta_i-z_i\alpha_c)^t\tau^{-1}(\theta
_i-z_i\alpha_c)\biggr\}.
\nonumber
\end{eqnarray}
Here $n_{i,c}$ is the number of $c_i$ that are qual to $c$, $c_{-i}$
are all the $c_j$ for $j\neq i$ and $b$ is such that if $c=c_j$, then
$\sum_{j\dvtx j\neq i} \{P(c_i=c)\}+P(c_i\neq c_j\ \forall j\neq i)=1$.
Next, we compute the probability that $c_i$ is different to any other
element in $\mathbf{c}$ as
%
\begin{eqnarray}\label{eqp2}
&&
P(c_i\neq c_j \mbox{ for all } j\neq i \mid c_{-i},\theta_i,\alpha)\nonumber\\
&&\qquad =
b\frac{M}{n-1+M}(2\pi)^{-p/2}|\tau|^{-1/2}|R|^{-1/2}|D_i|^{1/2}
\\
&&\qquad\quad{}\times\exp\biggl\{-\frac{1}{2}\bigl[\theta_i^t\tau^{-1}\theta_i-
[\theta_i^t\tau^{-1}z_i]D_i[z_i^t\tau^{-1}\theta_i]\bigr]\biggr\}.
\nonumber
\end{eqnarray}
If the imputed value of $c_i$, sampled based on (\ref{eqp1}) and
(\ref{eqp2}), is not associated with any other observation, it is
necessary to draw a value of $\alpha_{c_i}$ from~$H_i$, the
posterior distribution for $\alpha$ based on the prior $G_0$ and the
single observation $\theta_i$. In our case $H_i$ is given by $H_i\equiv
N_{pk}(\tilde{\alpha}_i, D_i)$, where
$D_i=[z_i^t\tau^{-1}z_i+R^{-1}]^{-1}$, and
$\tilde{\alpha}_i=D_i[z_i^t\tau^{-1}\theta_i]$.

\subsection*{Step 2}

In the second step, for all $c \in\{c_1,\ldots,c_n\}$ we draw a new
value~$\alpha_c$ given all the $\theta_i$ for which $c_i=c$, that is,
from the posterior distribution based on the prior $G_0$ and all the
data points currently associated with latent class~$c$. In our case,
this is given by $N_{pk}(\tilde{\alpha}_c,E)$, where
$E=[\sum_{i\dvtx c_i=c}z_i^t\tau^{-1}z_i+R^{-1}]^{-1}$ and
$\tilde{\alpha}_c=E[\sum_{i\dvtx c_i=c}z_i^t\tau^{-1}\theta_i]$.

Now we list all the full conditional distributions for the parametric
part of the model. The specific derivation details are straightforward
and therefore omitted.

\begin{itemize}
\item For fixed effect parameters we have
\[
\beta_j\mid\mbox{other parameters and data}\sim N_p(\tilde{\beta
_j},V_j),
\]
where
\[
\hspace*{-1.9pt}\tilde{\beta_j}=V_j\Biggl[\sum_{u=1}^{g}
\Biggl\{\Sigma_u^{-1}\sum_{i=1}^{n_u}\{
x_{ij}y_i-x_{ij}x_{il_1}\beta_{l_1}-\cdots-x_{ij}x_{il_q}\beta
_{l_q}-x_{ij}\theta_i\}\Biggr\}+\Lambda^{-1}\beta_{0j}\Biggr]
\]
and
\[
V_j=\Biggl[\sum_{u=1}^g\Biggl\{\sum_{i=1}^{n_u} x_{ij}^2\Sigma_u^{-1}\Biggr\}+\Lambda
^{-1}\Biggr]^{-1},
\]
where
\[
(l_1,l_2,\ldots, l_q)\neq
j,\qquad
j=1,\ldots,q.
\]
\item For the random effects parameters
$\theta_{1u},\ldots,\theta_{nu}$, $u=1,\ldots,g$, we have
that
\[
\theta_{iu}\mid\mbox{other parameters and data} \sim N_p(\tilde
{\theta}_{iu},Q_u), \qquad  i=1,\ldots,n,
\]
where
\[
Q_u=[\tau^{-1}+\Sigma_u^{-1}]^{-1}  \quad\mbox{and}\quad
\tilde{\theta}_{iu}=Q_u[\tau^{-1}z_i\alpha_i+\Sigma_u^{-1}y_i-\Sigma_u^{-1}Bx_i].
\]
\item For hyperparameters $\beta_{01},\ldots, \beta_{0q}$
we have
\[
\beta_{0j}\mid\mbox{other parameters and data} \sim N_p(\tilde{\beta}
_{0j}, D_0),
\]
where
\[
B_{0j}=D_0[\lambda^{-1}\beta_j+\tau_0^{-1}\beta_0], \qquad
j=1,\ldots,q,
\quad \mbox{and} \quad  D_0=[\Lambda^{-1}+\tau_{0}^{-1}]^{-1}.
\]
\item For hyperparameter $\Lambda$ we have
\[
\Lambda\mid\mbox{other parameters and data}\sim IW_p(d,E),
\]
where
\[
E=\sum_{j=1}^q(\beta_j-\beta_{0j})(\beta_j-\beta_{0j})^t+L_0
\quad\mbox{and} \quad  d=q+t_0.
\]
\item Finally, for the covariance matrices
$\Sigma_1,\ldots,\Sigma_g$, $\tau$ and $R$ we have
\[
\Sigma_u\mid\mbox{other parameters and data}\sim IW_p(l_u,H_u),
\]
where
\[
H_u=\sum_{i=1}^{n_u}(y_i-Bx_i-\theta_i)(y_i-Bx_i-\theta_i)^t+Q_0
\quad\mbox{and} \quad  l_u=n_u+\nu_0,
\]
\[
\tau\mid\mbox{other parameters and data}\sim IW_p(s,T),
\]
where
\[
T=\sum_{i=1}^{n}(\theta_i-z_i\alpha_i)(\theta_i-z_i\alpha_i)^T+\Phi_0
\quad \mbox{and} \quad  s=n+\gamma_0,
\]
\[
R\mid\mbox{other parameters and data}\sim IW_{pk}(f,O),
\]
where
\[
O=\sum_{i=1}^n\alpha_i\alpha_i^t+R_0  \quad\mbox{and} \quad  f=n+r_0.
\]
\end{itemize}
\end{appendix}

\section*{Acknowledgments}

The first author thanks the Comisi\'on Nacional de Investigaci\'on
Cient\'{\i}fica y Tecnol\'ogica---CONICYT, for supporting his Ph.D.
studies at the Pontificia Universidad Cat\'olica de Chile.



\printaddresses

\end{document}